\documentclass[prl,twocolumn]{revtex4}
\usepackage{graphicx}  
\usepackage{dcolumn}  
\usepackage{bm}          
\usepackage{amsmath}
\usepackage{amsfonts}
\usepackage{amssymb}

\begin{document}

\title{Conformal mapping of ultrasonic crystals: confining ultrasound
  and choclear-like wave guiding.}

\author{A. Baz\'an and M. Torres}
\affiliation{Instituto de F\'{\i}sica Aplicada, Consejo Superior de
  Investigaciones Cient\'{\i}ficas, Serrano 144, 28006 Madrid, Spain.}

\author{F.R. Montero de Espinosa}
\affiliation{Instituto de Ac\'ustica, Consejo Superior de
  Investigaciones Cient\'{\i}ficas, Serrano 144, 28006 Madrid, Spain.}

\author{R. Quintero-Torres and J.L. Arag\'on}
\affiliation{Centro de F\'{\i}sica Aplicada y Tecnolog\'{\i}a
  Avanzada, Universidad Nacional Aut\'onoma de M\'exico, Apartado
  Postal 1-1010, Quer\'etaro 76000, M\'exico.}

\begin{abstract}
  Conformal mapping of a slab of a two-dimensional ultrasonic crystal
  generate a closed geometrical arrangement of ultrasonic scatterers
  with appealing acoustic properties. This acoustic shell is able to
  confine ultrasonic modes. Some of these internal resonances can be
  induced from an external wave source. The mapping of a linear defect
  produces a wave-guide that exhibits a spatial-frequency selection
  analogous to that characteristic of a synthetic \textquotedblleft
  cochlea\textquotedblright ~. Both, experimental and theoretical
  results are reported here.
\end{abstract}


\maketitle

It has been recently reported that a parallel-sided slab of material
with negative refractive index can work as a perfect lens
\cite{Pendry1} and conformal transformations applied to these slabs
generate a variety of lenses \cite{Pendry2}. Also, quite recently,
curvilinear coordinate transformations and conformal mappings have
been applied to versatile optic metamaterials in order to develop
cloaks of invisibility \cite{Pendry3}. On the other hand, phononic
crystals, \emph{i.e.}, periodic composite materials whose structure
varies on the scale of the wavelength of the sound and exhibiting
forbidden frequency band-gaps, have also stirred a great interest
\cite{Sigalas,Sala,Montero,Vasseur,Psarobas,Liu,Yang1}.  Here, we
apply a conformal transformation to a slab of a conventional
two-dimensional ultrasonic crystal to build a continuous field of
ultrasonic scatterers that displays conspicuous physical
results. Several acoustic properties of this transformed struture are
studied theoretically and experimentally.

The analytic function using in this letter to generate the conformal
mapping is $w = \exp(z)$, where $w = u + iv$, and $z = x + iy$; $z$
belonging to the original complex plane and $w$ to the transformed one
\cite{Nehari}. This exponential function transforms parallel lines of
the original slab into concentric circumferences. Since this
continuous transformation preserves the angles between grid lines but
not the size of the objects, the transmission spectra is disturbed in a 
complex way, displaying a lot of localized resonant modes. The similarity 
ratio in the transformed arrangement is locally equal to $|dw/dz|$, preserving 
the filling ratio of the transformed unit cell and, hence, the effective medium
parameters for the long wavelength limit \cite{Berryman} are also
conserved. This property is characteristic of conformal structures but
is not fulfilled in the recently reported circular photonic crystals
\cite{Horiuchi}.

\begin{figure}[t!]
\begin{center}
\includegraphics[width=7.0cm]{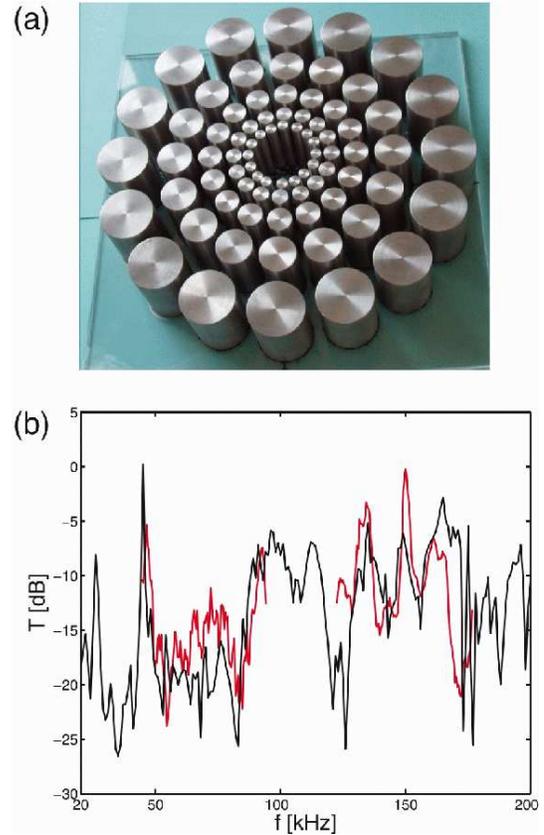}
\caption{(a) The conformal radial arrangement of aluminum ultrasound
  scatterers.(b) Experimental (red) and theoretical (black)
  transmission spectra trough four layers of the conformal device. The
  radius of the inner circumference is 22.5 mm and the diameter of its
  corresponding cylinders is 7.5 mm. The radius of the outer
  circumference is 113.9 mm and the diameter of its cylinders is 37.9
  mm.}
 \label{fig:fig1}
\end{center}
\end{figure}

By applying the exponential conformal mapping to an adequate square
periodic slab of identical circles, we generate the radial structure
shown in Fig.\ref{fig:fig1}(a). Aluminum cylinders were glued to a
methacrylate layer with 6 mm of thickness. Cylinders, with a height of
5 cm, were turned from bars of aluminum UNE 6082 (L 3453). The Young
modulus of this material is $E$ = 6.9x10$^{10}$ N/m$^2$ and its
density is $\rho$ = 2.7x10$^3$ Kg/m$^3$. These cylindrical scatterers
were immersed in a large vessel, provided with anechoic walls, and
filled with degasified water with a depth of 6 cm.

To measure the acoustic transmission through this shell of immersed
cylindrical scatterers we emit with a Langevin type piezoelectric
sandwich, built by using two PZT-5A piezoelectric rings. Te diameter
of the is of $D$=15.5 mm and the length, without coating, of $L$ = 33
mm. The launched signal is always received in a PVDF needle hydrophone
DAPCO NP 10-3A90 and processed by means of a Hewlett Packard 4194A
Impedance/Gain-Phase Analyzer. Theoretical calculations are carried
out by solving the acoustic propagation equation by means of the
Finite Element Method (FEM). We only present experimental measurements
within the two broad resonance bands of the piezoelectric sandwich,
ranging from 50 to 100 kHz and from 120 to 160 kHz respectively.

In Fig.\ref{fig:fig1}(b) we show both theoretical and experimental
measurements of the outgoing transmission spectrum through the
conformally mapped structure. The transmitter is now the transducer
diametrically placed near the inner circular layer of scatterers and
the receiver is the needle hydrophone located just at the output of
the structure at the end of the corresponding emission radius. This
measurement scheme is always chosen in order to profit the high
directionality of the sandwich transducer, thus overcoming the high
energy damping of the system.

\begin{figure}[t!]
\begin{center}
\includegraphics[width=7.0cm]{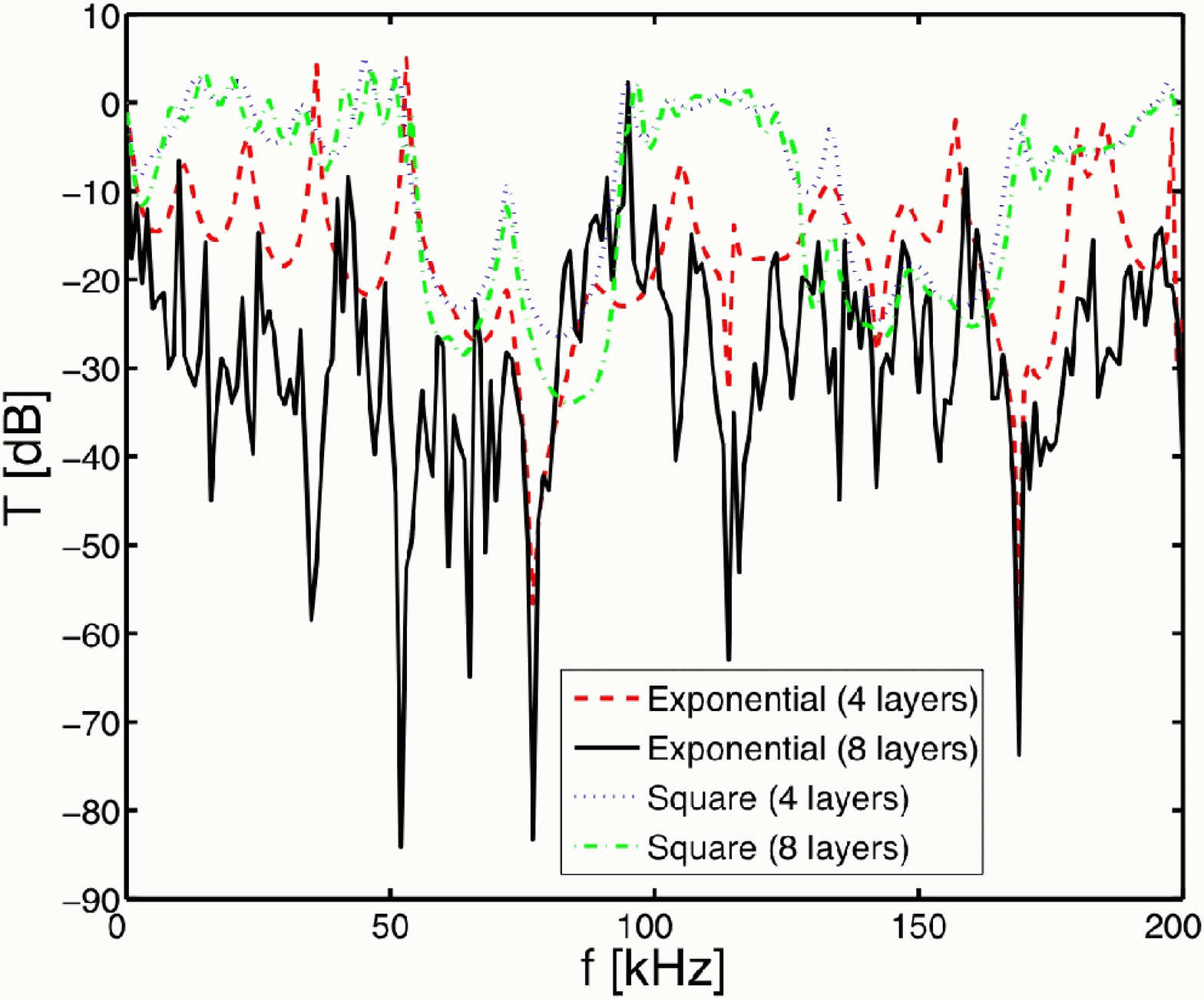}
\caption{Theoretical transmission spectra trough the original slab
  containing four (blue) and eight (green) layers of scatterers
  respectively. Transmission spectra trough the conformal structures
  containing four (red) and eight (black) circular layers of
  ultrasound scatterers, respectivelly.}
 \label{fig:fig2}
\end{center}
\end{figure}

\begin{figure}[t!]
\begin{center}
\includegraphics[width=7.0cm]{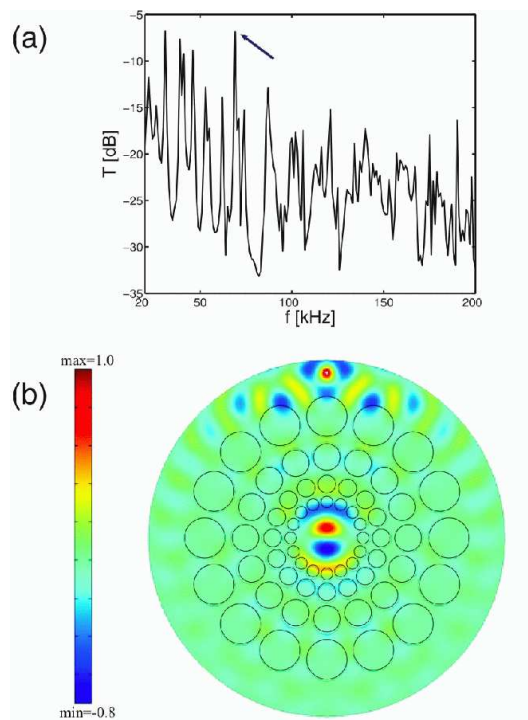}
\caption{(a) Ingoing transmission spectra for the conformal
  arrangement of ultrasound scatterers; the arrow points out the inner
  resonance shown in (b), where it is injected from the external
  boundary at 69 kHz.}
 \label{fig:fig3}
\end{center}
\end{figure}

In Fig.\ref{fig:fig2}, transmission spectra FEM calculations for both
the original square and its corresponding conformally mapped crystal
slab are compared. Calculations are presented for slabs containing
four and eight rows of scatterers respectively. The original
ultrasonic crystals are excited by means of plane waves parallel to
the slabs and transmission spectra are calculated by integrating the
received signal after passing through the slabs. In the conformal
structures the exciting waves are concentric circular waves launched
from the inner of the cavities and their corresponding transmission
spectra are calculated by integrating the received signal along a
concentric circle enclosing the structures. The transmission spectra
of the original slabs exhibit well defined forbidden frequency band
gaps, as expected. On the contrary, the corresponding spectra for the
transformed structures exhibit a spiky structure showing many deep
peaks corresponding to internal resonances. The more of the layers in
the circular shell the more peaks in the spectrum. Thus, these
conformal structures can be used as ultrasonic traps to confine
internal modes with a wide and discrete range of frequencies and as
ultrasonic filters exhibiting very high attenuation for some frequency
ranges. The present conformal device also allows to implement the
important technique that consists of injecting inner resonances
into the cavities of the shells starting from external point wave
sources. Fig.\ref{fig:fig3} (a) shows the ingoing transmission
spectra for the conformal structure. The transmitter source is placed
near the external boundary of the scatterer arrangement and the
integrating receiver circular domain is located just at the center of
the cavity. We have found a rich variety of inner localized modes
injected from the external boundary. In Fig.\ref{fig:fig3}(b) we show
an example of the injection of an internal resonance in the conformal
shell.

\begin{figure}[t!]
\begin{center}
\includegraphics[width=8.0cm]{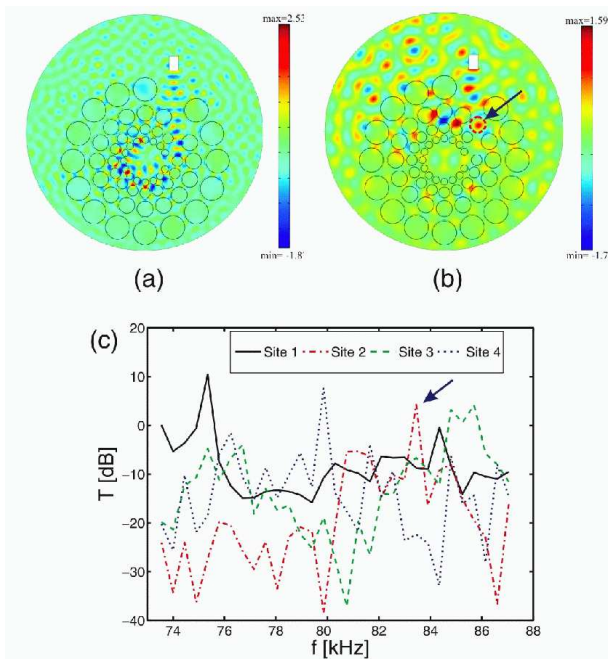}
\caption{(a) Wave guiding along a spiral-like defect excited at 148
  kHz trough the conformally mapped device. (b) Experimental
  measurements of the spatial-frequency discrimination performed at
  the four defects, from outer (1) to inner (4), conforming the
  choclear-like guide. The red arrow points out a resonance measured
  at 83 kHz and located at the second defect. (c) Theoretical
  calculation showing that this resonance at 83 kHz is predicted.}
 \label{fig:fig4}
\end{center}
\end{figure}

By removing a diagonal line along the $[11]$ direction in the
original rectangular slab, a spiral-like defect is generated in the
conformally mapped structure. In Fig.\ref{fig:fig4}(a) we show the
guiding of the wave along these linear defects. There are a lot of
resonance frequencies to localize the wave along these spiral
choclear-like defects. In Fig.\ref{fig:fig4}(b) we show the
experimental measurements performed at four different locations along
the choclear-like defect generated in the conformal structure, where
the numerical label of each curve indicates the relative location
along the defect path, from the exterior to the interior. As it can be
seen in the figure, a clear spatial- frequency selection is
achieved. This performance allows us to identify this wave guide with
a device similar to a \textquotedblleft synthetic
choclea\textquotedblright, which is able to discriminate and separate
different frequencies by exploring the path along the defect with
different point receptors for each resonance. So, the spiral-like
defects works as the \textquotedblleft
Negative-print\textquotedblright ~ or \textquotedblleft
mould\textquotedblright ~ of a natural choclea. As an example, in
Fig.\ref{fig:fig4}(c) we show the theoretically FEM calculated
resonance of this choclear-like defect corresponding to a frequency of
83 kHz. As it can be seen, the theoretical prediction, indicated with
a little arrow in Fig.\ref{fig:fig4}(b), fits with high accuracy a
remarkable peak of the experimental observation shown in the second
curve in Fig.\ref{fig:fig4} (c). Theoretical and experimental
concordance is observed also for the other resonance peaks.

As conclusion, a conformal mapping of ultrasonic crystals give rise to
a variety of interesting acoustic transmission phenomena.

\begin{acknowledgments}
  This work has been partially supported by the Spanish MCYT (Grant
  No. FIS2004-03237), and the Mexican DGAPA-UNAM (Grant Nos. IN-117806
  and IN-118406) and CONACyT (Grant No. D40615-F).
\end{acknowledgments}

\end{document}